\documentclass[3p,times]{elsarticle}
\usepackage{amssymb}

\usepackage[figuresright]{rotating}
\usepackage{xcolor}
\usepackage{color,soul}
\usepackage{amsmath}
\usepackage{siunitx}
\usepackage{hyperref}
\usepackage{multirow}

\begin{document}

\begin{frontmatter}




\title{Functionalized carbophenes as high-capacity versatile gas adsorbents: An ab initio study}


\author[label1,label2]{Chad E. Junkermeier}
\author[label3]{Evan Larmand}
\author[label3]{Jean-Charles Morais}
\author[label2]{Jedediah Kobebel}
\author[label1]{Kat Lavarez}
\author[label1]{R. Martin Adra}
\author[label1]{Jirui Yang}
\author[label1]{Valeria Aparicio Diaz}
\author[label4]{Ricardo Paupitz}
\author[label3]{George Psofogiannakis}

\address[label1]{Department of Physics and Astronomy, University of Hawai`i at Mānoa, Honolulu HI 96822, USA}
\address[label2]{Science, Technology, Engineering, and Mathematics Department, University of Hawai`i Maui College, Kahului HI 96732, USA}
\address[label3]{Department of Chemical and Biological Engineering, University of Ottawa, Ottawa, Ontario, Canada}
\address[label4]{Departamento de F\'{\i}sica, IGCE, Universidade Estadual Paulista, UNESP, 13506-900, Rio Claro, SP, Brazil}

\begin{abstract}

This study employs density functional theory (DFT) and density functional tight-binding theory (DFTB) to determine the adsorption properties of carbon dioxide (CO$_2$), methane (CH$_4$), and dihydrogen (H$_2$) in carbophenes functionalized with carboxyl (COOH), amine (NH$_2$), nitro (NO$_2$), and hydroxyl (OH) groups. We demonstrate that carbophenes are promising candidates as adsorbents for these gasses. Carbophenes have larger CO$_2$ and CH$_4$ adsorption energies than other next-generation solid-state capture materials. Yet, the low predicted desorption temperatures mean they can be beneficial as air scrubbers in confined spaces. Functionalized carbophenes have H$_2$ adsorption energies usually observed in metal-containing materials. Further, the predicted desorption temperatures of H$_2$ from carbophenes lie within the DOE Technical Targets for Onboard Hydrogen Storage for Light-Duty Vehicles (DOEHST) operating temperature range. The possibility of tailoring the degree of functionalization in combination with selecting sufficiently open carbophene structures that allow for multiple strong interactions without steric hindrance (crowding) effects, added to the multiplicity of possible functional groups alone or in combination, suggests that these very light materials can be ideal adsorbates for many gases. Tailoring the design to specific adsorption or separation needs would require extensive combinatorial investigations.



\end{abstract}


\end{frontmatter}

\section*{Introduction}

The emission of greenhouse gases (GHG) from human activities is the most significant cause of climate change, manifested by an increase in global mean surface air temperatures over the last 100 years. Climate change is leading to disastrous consequences such as increasingly extreme weather and climate events, thus necessitating a global effort to reduce the release of GHG into the atmosphere~\cite{stocker2014climate}. Carbon capture technology, such as membrane gas separation, is a potential part of the climate change solution. For example, gas capture systems can be operated continuously, with GHG separated from exhaust gas streams of processes~\cite{Zhao17}.

In addition, spacecraft, submarines, and mining operations utilize various materials to capture carbon dioxide to maintain a breathable atmosphere in tight spaces~\cite{winton2016carbon}. For example, lithium hydroxide (LiOH) is extensively used to remove carbon dioxide from the air as a retention adsorbent~\cite{Zilberman20154}. LiOH canisters act as chemical adsorbents, reacting with carbon dioxide to form lithium carbonate and water. However, LiOH canisters have limited reusability and produce waste products that must be disposed of or recycled. Solid amine sorbents, such as activated carbon or zeolites, have shown promise as alternative capture materials. These materials possess high surface areas and can physically or chemically adsorb carbon dioxide. In addition, solid amine sorbents are reusable, potentially reducing the waste generated during space missions. For example, NASA is readying a new generation of CO$_2$ scrubber for the International Space Station, which employs the zeolite sorbent 13X in the 4BMS CO$_2$ Scrubber~\cite{cmarik2018co}.

Covalent Organic Frameworks (COFs) have aroused much interest for use in carbon capture and gas separation. The high porosity, highly ordered structures, and well-organized nanopores led to these structures being identified as exciting candidates for applications in gas separation, among many other possible applications~\cite{D0CS00049C}.

DFT has previously been used to determine the adsorption energies of functional groups and GHG adsorbates on functional groups in COFs. For example, Lim \textit{et al.} evaluated the adsorption of CO$_2$ in carbonaceous materials with various functional groups for carbon capture applications~\cite{Lim5b12090}. In some models, the COFs were approximated with aromatic molecules. For example, Arjmandi \textit{et al.} used a benzene dicarboxylate base structure and applied DFT to obtain the adsorption energy for CO$_2$ adsorbate on multiple functional groups~\cite{Arjmandi20181473}. Similarly, Torrisi \textit{et al.} worked with a benzene base structure with CO$_2$ as an adsorbate~\cite{Torrisi20093120909, Torrisi2010044705}. In addition, periodic framework models of materials have been tested for adsorption properties using DFT, as seen in the study by Soleymanabadi and Kakemam, who studied functionalized carbon nanotube structures for H$_2$ adsorption~\cite{SOLEYMANABADI2013115}.

Solid state materials are also investigated for hydrogen storage~\cite{HIRSCHER2020153548, SIMANULLANG202229808}.
Because dihydrogen has the highest energy density of any fuel, research into its uses as an energy source in everything from cell phones to motor vehicles is ongoing. The structure of the solid-state material largely determines H$_2$ adsorption energies, examples of which include van der Waals materials (physisorption, $ <$0.1 eV), metal-organic frameworks (enhanced physisorption, 0.1-0.2 eV), metal catalysts on low dimensional materials (Kubas binding, 0.2-0.3 eV), noble metal catalysts on reducible metal oxides (spillover effect, $<$0.5 eV), metal hydrides (chemisorption, $\sim$0.5-1.0 eV), chemical hydrides (chemisorption, $>$1.0 eV)~\cite{dillon2005nrel, D2SC00871H}. Despite the intense effort, storage options that meet the DOEHST have not yet been produced~\cite{us2017doehydrogen}. 

In this study, we will consider the use of N-carbophenes as a possible class of materials for use in fighting climate change by adsorbing the greenhouse gases CO$_2$ and CH$_4$ or for use in alternative fuel (i.e., H$_2$) storage. The structures used in this study were adapted from a structure Du \textit{et al.} proposed as a possible material produced when trying to synthesize graphenylene~\cite{du1740796}. Based on the relative formation energies of graphenylene and 3-carbophene, Junkermeier \textit{et al.} concluded that 3-carbophene might have been synthesized by Du \textit{et al.}~\cite{junkermeier2019simplecarbophene}. They also demonstrated that the interlayer spacings of 3-carbophenes are closer to Du's experimentally produced materials than graphenylene's interlayer spacing. 3-carbophene is one example of a novel class of two-dimensional covalent organic frameworks (2DCOFs) called N-carbophenes. N-carbophenes have hexagonal pores with pore edges of alternating N cyclohexatriene and N-1 cyclobutene units. Later, Junkermeier \textit{et al.} demonstrated that functionalized carbophenes are even more energetically favorable than graphenylene~\cite{Junkermeier2022covalent}.

The objectives of this study were several-fold. Firstly, to evaluate the gas adsorption properties (i.e., adsorption energy) of pristine carbophenes using the gases CO$_2$, CH$_4$, and H$_2$. Second, to compute the adsorption energies of carbophenes modified by the presence of the functional groups often found after synthesizing 2D carbons (i.e., COOH, NH$_2$, NO$_2$, OH). Third, to determine the effect of having multiple functional groups in a pore on adsorption energy. Lastly, the material properties, including molecular weight and density, were to be determined. In this work, we've demonstrated that the adsorption energies of GHG adsorbates into functionalized carbophenes are greater than proposed next-generation adsorbents while being half as dense. Adding a second GHG molecule does not significantly change the adsorption properties, though having a higher density of functional groups within the carbophene pores can decrease the adsorption energy of the GHGs. It has also been demonstrated that functionalized carbophenes can adsorb H$_2$ molecules with energies rivaling metal-containing materials. The predicted desorption temperatures are $> 400$ K, meaning carbophenes may be used as CO$_2$ as air scrubbers in confined spaces or solid-state H$_2$ storage.

\section*{Methodology}\label{sec:methods}

Density functional theory (DFT) calculations were used to determine the adsorption properties of functionalized 3-carbophene, while density functional-based tight-binding (DFTB) calculations were used to extend to larger systems.

The DFT calculations in this study were performed in ADF-BAND (Version ADF.2019) using the Triple Zeta plus Polarization (TZP) basis set, GGA:PBEsol-D3(Bj) exchange-correlation functional, and Grimme D3 dispersion correction~\cite{Perdew2008136406, PhysRevB.44.7888, GrimmeD3}.  All other parameters were set to the standard settings of the Modeling Suite (AMS) graphic user interface. We used the default options generating the automatic regular k-space grid for quality “normal” consisting of a single k-point.

The DFTB calculations were performed using DFTB+ (pre-compiled Version 19.1)~\cite{Elstner1998,aradi2007dftb,manzano2012}. DFTB+ has near DFT precision in electronic structure calculations while significantly faster than DFT. Much of this speedup comes from using look-up tables (Slater-Koster files) instead of integral evaluation at run-time.
The {\sl matsci Slater-Koster} files, formulated to describe materials science problems accurately, were used~\cite{frenzel2004semi, lukose2010reticular}. The Grimme D3 formulation was used to simulate dispersion forces~\cite{GrimmeD3}. Hydrogen bonding terms were corrected using the (HCorrection = H5) method~\cite{Rezac20174804}.
Geometry optimization, including cell parameters and atomic positions, used a conjugate gradient algorithm with a maximum force difference of 10$^{-5}$ Ha/Bohr and an SCC maximum tolerance of 10$^{-4}$ electrons as convergence criteria. The lattice vector lengths were allowed to change during geometry optimization, but the angle between them was not. An 8x8x1 Monkhorst-Pack kpoint grid was used during optimization. The out-of-plane bounding box was set to 30 \AA.  A sample job script that includes the DFTB+ input file parameters used in this work can be found in Reference \cite{gasstoragecarbophenes}.

\section*{Results and Discussion}

\begin{figure*}
\centering
\includegraphics[clip,width=6 in, keepaspectratio]{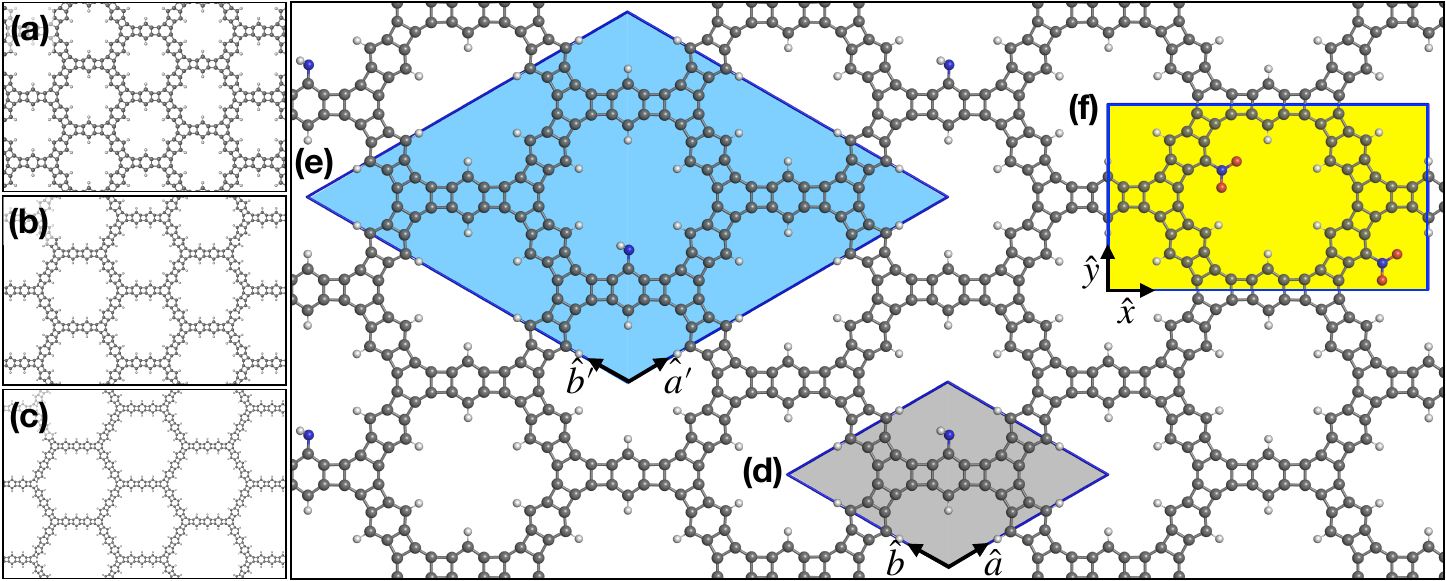}
\caption{Examples of the pristine (a) 3-, (b) 4-, and (c) 5-carbophene used in this work. Examples of functionalized 3-carbophene: (d) Primitive unit cell, (e) 2-by-2 supercell used in DFTB+ calculations, and (f) rectangular supercell used with ADF-BAND calculations.}
\label{fig:LatVecSupercell}
\end{figure*}

This work focuses on carbon capture in functionalized 3-, 4-, and 5-carbophenes, examples of which are shown in Figure~\ref{fig:LatVecSupercell}. Starting from the lattice parameters given by Junkermeier \textit{et al.}, two types of pristine carbophene supercells were produced 1) a 2-by-2 supercell designed to increase the distance between functional groups such that they are unlikely to be interacting, 2) a rectangular supercell designed to minimize the number of atoms in a cell structure while providing a complete pore~\cite{junkermeier2019simplecarbophene, Solenov13115502}. Functional groups (i.e., NH$_2$, OH, COOH, NO$_2$, F) were added to the optimized pristine carbophenes, and the cell structure and atomic positions were then optimized. Figure~\ref{fig:LatVecSupercell} (d-f) shows the unit lattice vectors (black arrows) and bounding boxes (blue lines) of (a) 3-carbophene primitive unit cell (light gray parallelogram), (b) 3-carbophene 2-by-2 supercell (light blue parallelogram), and (c) 3-carbophene rectangular supercell (yellow rectangle). Similar supercells were produced for 4- and 5-carbophenes. In Figure~\ref{fig:LatVecSupercell}, the image is based on the functionalization pattern used in the 2-by-2 supercells with the DFTB+ calculations. In contrast, the yellow rectangular supercell is used in the ADF-BAND DFT calculations with one functional group per pore. The rectangular supercell is also used in ADF-BAND calculations with multiple functional groups per pore.

Table \ref{table:chadtable1} presents the lattice constants and molecular weights of single-layer functionalized 3-carbophene.  Assuming the interlayer spacing of 3.54 Å for bulk carbophenes found in~\cite{junkermeier2019simplecarbophene}, Table \ref{table:chadtable1} also presents the densities of the functionalized bulk 3-carbophenes.  Due to their large pore diameters (i.e., $>$ 7.5 Å), the densities of the bulk 3-carbophenes are about half of typical zeolites values (i.e., 2.2 - 2.5 g/cm$^3$).  Metal-organic frameworks (MOFs) may have densities less than 0.15 g/cm$^3$ but also depend on expensive metal atoms~\cite{Li2017624}.  But if we suppose that bulk 5-carbophenes can be synthesized, the density of bulk 5-carbophene drops to 0.61 g/cm$^3$, which is on par with the most studied MOFs.  Activated carbons have densities of 0.4-0.5 g/cm$^3$ and are inexpensive.


\begin{table*}[!htb]
    \centering
\caption{Comparison of ADF-BAND and DFTB+ optimized, lattice constants (x,y,a$^\prime$) [Å], molecular weights (MW) [g/mol], and densities [g/cm$^3$] of pristine and functionalized 3-carbophenes.}
\begin{tabular}{cccccccc}
    \hline
Functionalization    &    x$_{\text{BAND}}$     &    y$_{\text{BAND}}$    &    a$^\prime_{\text{DFTB+}}$    &    MW$_{\text{BAND}}$     &    MW$_{\text{DFTB+}}$     &    $\rho_{\text{BAND}}$    &    $\rho_{\text{DFTB+}}$     \\
    \hline
 Pristine     &    23.03    &    13.32    &    26.94    &    732.74    &    1465.47    &    1.12    &    1.10    \\
 COOH     &    23.03    &    13.32    &    26.94    &    820.76    &    1509.48    &    1.26    &    1.13    \\
 NH2     &    23.03    &    13.32    &    26.94    &    762.77    &    1480.49    &    1.17    &    1.11    \\
 NO2     &    23.03    &    13.32    &    26.94    &    822.73    &    1510.47    &    1.26    &    1.13    \\
 OH     &    23.03    &    13.32    &    26.95    &    764.74    &    1481.47    &    1.17    &    1.11    \\
\hline
    \end{tabular}
    \label{table:chadtable1}
\end{table*}

\begin{figure}
\centering
\includegraphics[clip,width=3 in, keepaspectratio]{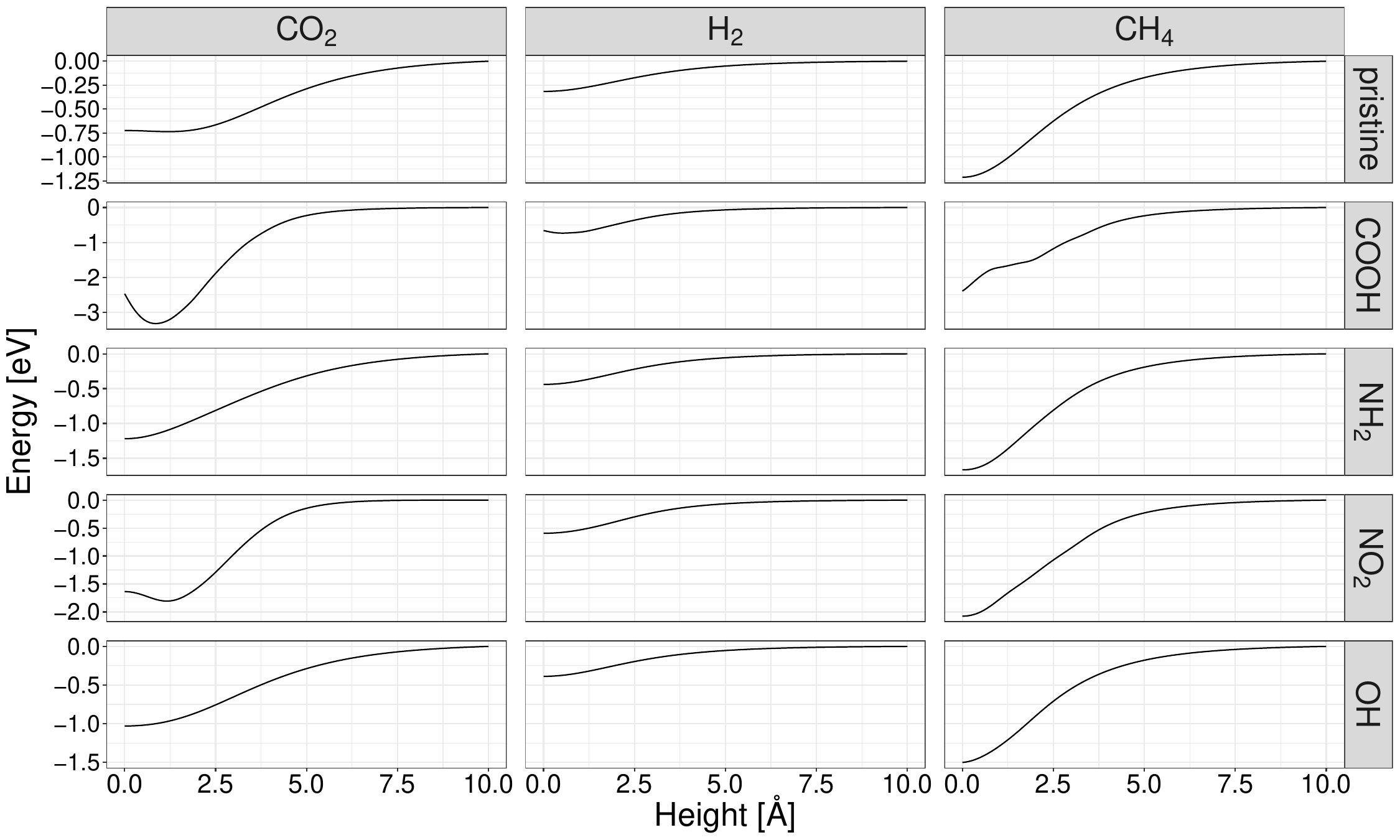}
\caption{Diffusion curves of gas molecules through the center of functionalized carbophene pores.}
\label{fig:barriers}
\end{figure}

Using DFTB+ single point calculations within a high throughput framework, we estimated the energy curves of CO$_2$, CH$_4$, and H$_2$ as they diffuse through the center of a 3-carbophene pore; see Figure~\ref{fig:barriers}. The single-point calculations used a position step increase of 0.025 Å from 0 to 4 Å and 0.1 Å from 4.1 to 10 Å. The results demonstrate that the functionalized carbophenes have deeper potential wells than pristine carbophenes. The COOH functionalized carbophene has the deepest potential well because the COOH functional has greater physisorption capabilities due to the lone pairs on the oxygen atoms and the possibility of hydrogen bonding due to the -OH. In most cases, the gas molecules have the lowest energy when they are at the center of the pore. For example, the polar CO$_2$ molecule has a minimum total energy at 0.85 Å with an energy barrier of 0.85 eV when diffusing through a COOH-functionalized pore. The COOH functionalization also produces a barrier of 0.075 eV for H$_2$ diffusion with a minimum energy at 0.525 Å from the pore. The trends found with these single-point calculation-based energy curves are borne out when the atoms in the gas molecule-functionalized carbophene systems relax.

Using the plane wave approximation of ADF-BAND, a computationally appropriate method for computing the adsorption energy ($E_{ads}$) is to find the energy difference between two relaxation calculations.  The first relaxation calculation has the carbophene and gas molecule in the same periodic box but well separated. A separation of 20 Å between the gas molecule and carbophene was determined to be sufficiently large.  The second relaxation calculation places the gas molecule within the carbophene pore. The $E_{ads}$ is calculated using the equation
\begin{equation} \label{eq:EadsBANDS}
E_{ads}=E_{dist} - E_{prox},
\end{equation}
where $E_{dist}$ is the total energy of the system where the gas molecule is far from the carbophene, so there are no interactions, and $E_{prox}$ is the total energy of a relaxed functionalized carbophene and adsorbed gas molecule system. Within the local orbital approximation used in DFTB+, Equation \ref{eq:EadsBANDS} is more naturally rewritten as
\begin{equation} \label{eq:EadsDFTB}
E_{ads}=E_{carbophene} + E_{gas} - E_{prox} ,
\end{equation}
where $E_{carbophene}$ is the total energy of the functionalized carbophene alone in the periodic cell, and $E_{gas}$ is the total energy of the gas molecule alone in the periodic cell. In Eqns.~\ref{eq:EadsBANDS} and \ref{eq:EadsDFTB}, the terms are ordered such that positive $E_{ads}$ indicates stable adsorption.

\begin{figure*}
\centering
\includegraphics[clip,width=1 \textwidth, keepaspectratio]{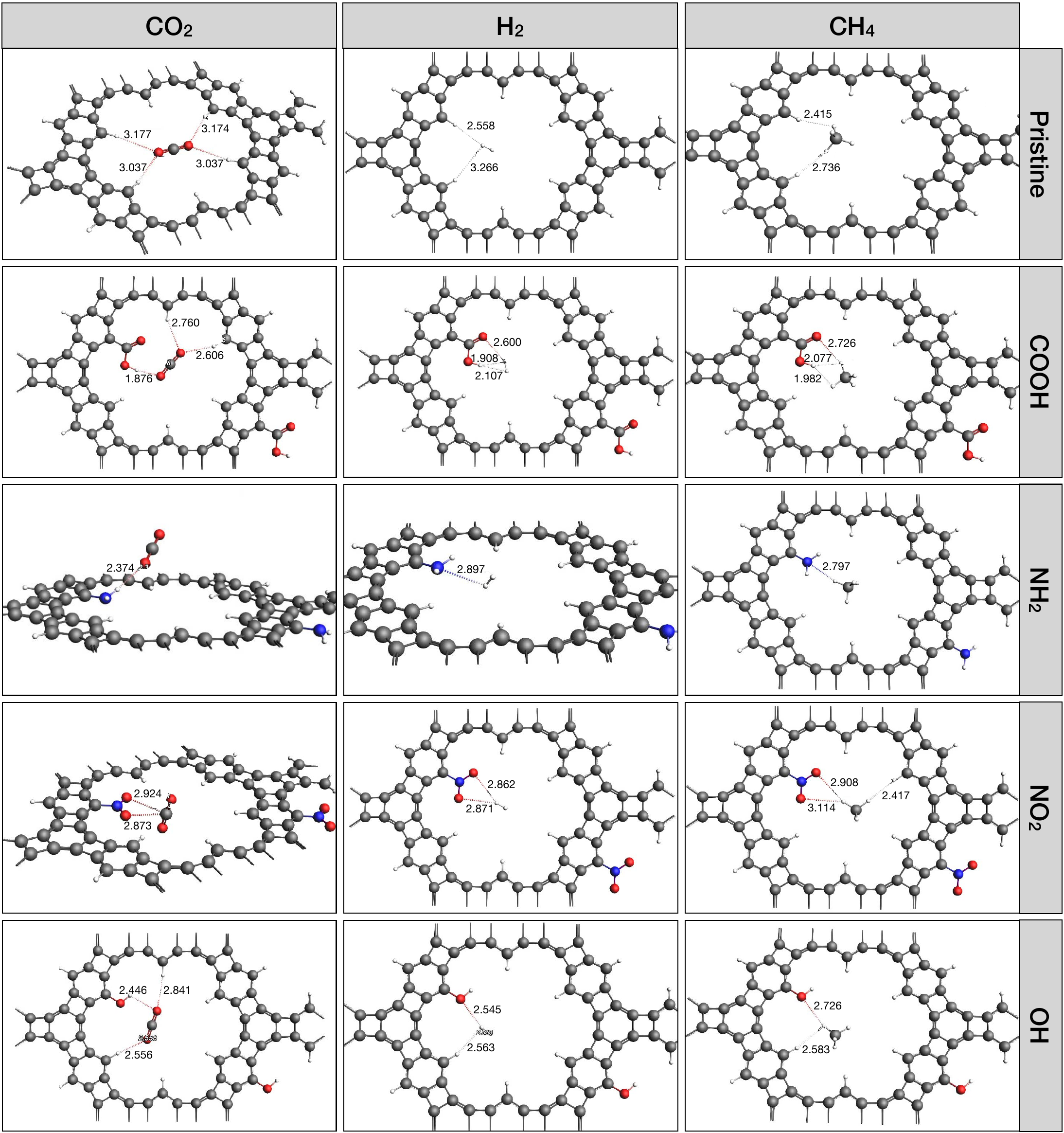}
\caption{DFT optimized gas molecule adsorption where the gas molecule was originally placed at the center of the carbophene pore. Dotted lines represent possible H-bonds and Coulomb interactions.}
\label{fig:EVANAppendixC}
\end{figure*}

\begin{figure}
\centering
\includegraphics[clip,width=3 in, keepaspectratio]{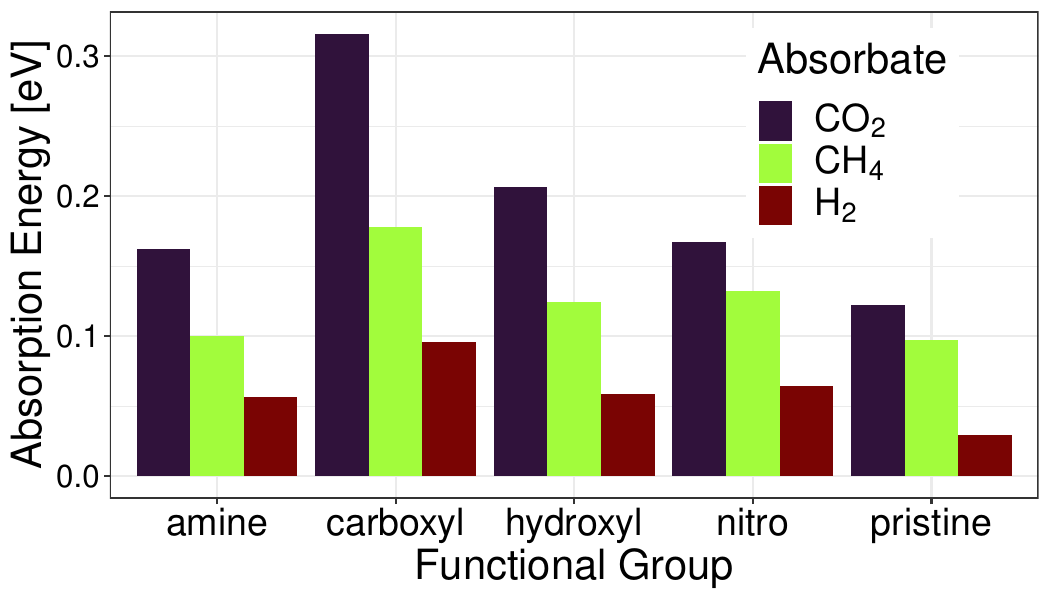}
\caption{Comparison of adsorption energies ($E_{ads}$), in eV, of the simulation results.}
\label{fig:EvanFigure1}
\end{figure}

We utilized DFT to investigate single gas molecules in rectangular supercells of functionalized 3-carbophene. In doing so, a gas molecule was placed in the center of a carbophene pore and relaxed. In each case, the gas molecule was near the center of the pore after optimization, as shown in Figure~\ref{fig:EVANAppendixC}.  Figure~\ref{fig:EvanFigure1} shows a graph of the adsorption energies. In particular, the results suggest that carboxyl functionalized 3-carbophene has the highest adsorption energy for all three adsorbents with adsorption energies of 0.32, 0.18, and 0.10 eV for CO$_2$, CH$_4$, and H$_2$, respectively.

The significant adsorption energies of carboxyl-functionalized 3-carbophene are likely a result of COOH having the highest number of possible molecular interactions compared to the other functional groups. Figure~\ref{fig:EVANAppendixC} shows CO$_2$ interacting with the adjacent -H atoms encircling the pore (at distances of 2.760 and 2.606 Å) in addition to the -H atom of the carboxyl group (1.676 Å). A Coulomb interaction exists between the lone pair on the double-bonded carbonyl -O atom and the CO$_2$ molecule. An adjacent -H atom may stabilize the carboxyl functional as it is also within hydrogen bonding distance. The H-bonding between the carboxyl group and an adjacent functional -H atom would increase the calculated adsorption energy while not strengthening the bonding situation of the CO$_2$ molecule. The double-bonded oxygen in the carboxyl group plays a significant role in the interactions with both H$_2$ (2.699 Å) and CH$_4$ (2.726 Å).

Regarding CO$_2$ adsorption, the next strongest interaction was with the hydroxyl-containing system (0.21 eV), which is likely due to CO$_2$ -O atoms' interactions with the -H atom of the hydroxyl group (2.446 Å) and the adjacent -H atom around the ring (2.556 and 2.841 Å). The nitro-CO$_2$ system was the next highest for CO$_2$ adsorption (0.16 eV), which is lower than the hydroxyl system since the vertical orientation of the CO$_2$ adopted does not allow interactions with the pore -H atoms that were available in the hydroxyl system. In comparison, two O-C interactions are found between the nitro group and CO$_2$ (2.924 and 2.873 Å). The amino-functionalized system had the next highest CO$_2$ adsorption energy, 0.06 eV. CO$_2$ had only one clear interaction with the amino group. The CO$_2$ hydrogen bonded with the amino -H (2.374 Å).  Finally, the pristine structure had the lowest CO$_2$ adsorption energy (0.12 eV). The CO$_2$ pristine carbophene system had four interactions between CO$_2$ oxygen atoms and the pore -H atoms (3.177, 3.174, 3.037, 3.037 Å). Though they are weak, they add up to significant interactions.

The nitro-containing system was the next highest for CH$_4$ and H$_2$ adsorption (0.13 and 0.06 eV, respectively), which is likely due to the two -O atoms from the nitro group interacting with CH$_4$ (2.906 and 3.714 Å) and H$_2$ (2.862 and 2.871 Å). Nitro’s unique structure always grants one of the two oxygen atoms a negative charge, which ensures at least one hydrogen binding site is available when interacting with hydrogen-containing molecules. CH$_4$ may also form an interaction with a distal -H atom (2.417 Å) in the nitro-CH$_4$ system via the Van der Waals force. CH$_4$ and H$_2$ had their next highest adsorption energies in the hydroxyl system (0.12 and 0.06 eV, respectively). The interactions between them are similar such that they both interact with the hydroxyl -H atom (2.726 and 2.545 Å for CH$_4$ and H$_2$ respectively) and the lower adjacent pore -H atom (2.583 and 2.563 Å for CH$_4$ and H$_2$ respectively) via van der Waals force. Additionally, depending on the reaction condition, a potential partial bond between an -O atom from the hydroxyl functional group and a -H atom from CH$_4$ or H$_2$ is possible.

The amino system had the next highest adsorption energies (0.10 and 0.06 eV for CH$_4$ and H$_2$, respectively). Each adsorbent had only one clear interaction with the amino group (2.797 and 2.897 Å for CH$_4$ and H$_2$, respectively). CH$_4$ and H$_2$ seem to line up to access the amino -N atom lone pair via hydrogen bonding. Finally, the pristine structures had the lowest interaction energies for CH$_4$ and H$_2$. With only Van der Waals interaction present, the orientation of CH$_4$ in the pristine structure allows for two interactions with the pore -H atoms (2.415 and 2.736 Å). Similarly, the H$_2$ system also interacted with the pore -H atoms (2.558 and 3.226 Å).


\begin{figure}
\centering
\includegraphics[clip,width=3 in, keepaspectratio]{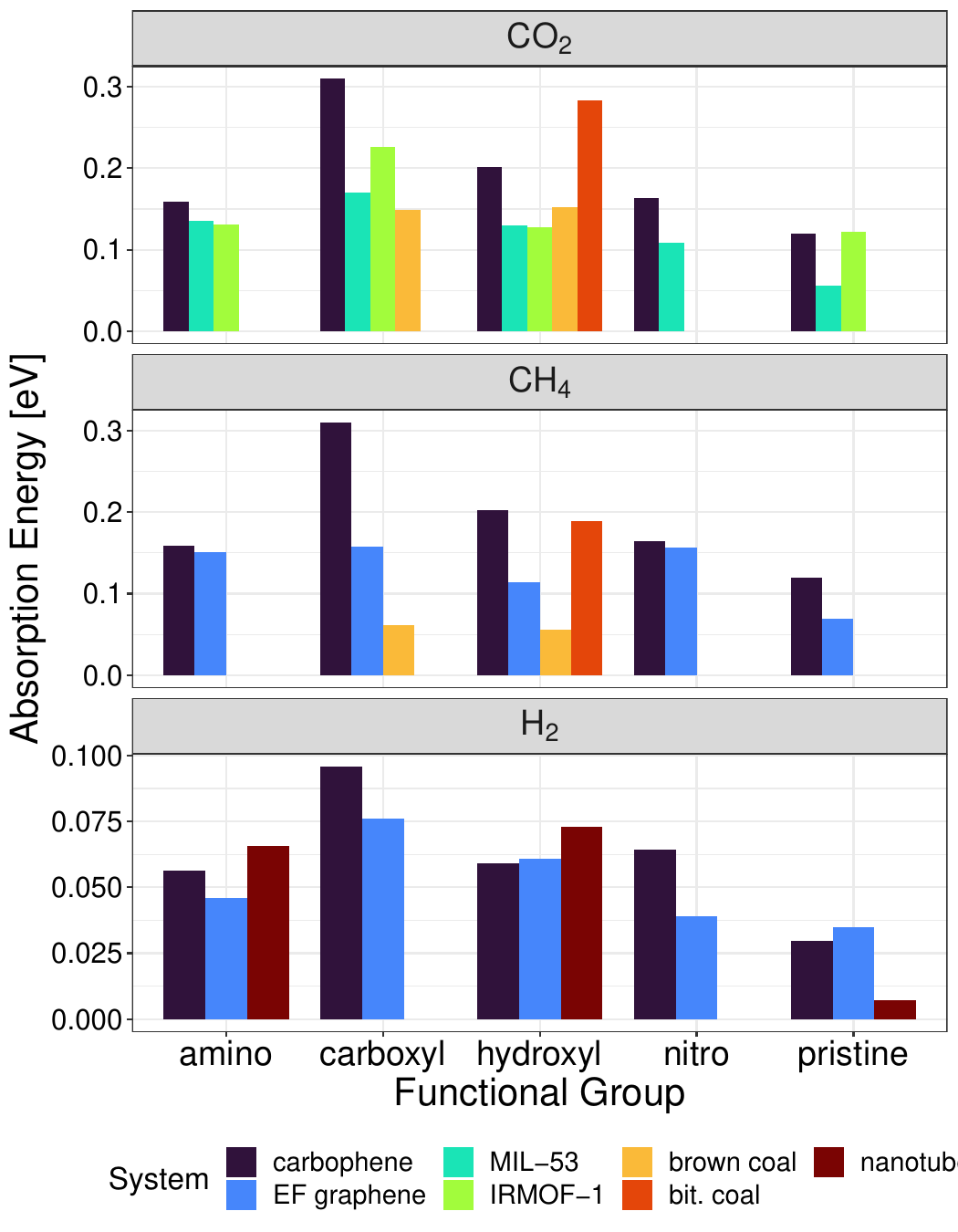}
\caption{Comparison of adsorption energies of CO$_2$, CH$_4$, or H$_2$ into functionalized 3-carbophene with values reported elsewhere.}
\label{fig:Evancomparison}
\end{figure}

These results can be compared to other similarly functionalized materials to gauge the relative strength of 3-carbophene as an adsorption material; see Figure~\ref{fig:Evancomparison}. Their authors considered each of these materials a possible replacement for the current generation of carbon-adsorbing materials. Figure~\ref{fig:Evancomparison} demonstrates that the 3-carbophene systems have larger adsorption energies in nearly all cases. The CO$_2$ adsorption in carbophenes has greater adsorption energies than the similarly studied metal-organic frameworks IRMOF-1~\cite{Arjmandi20181473}, MIL-53~\cite{Torrisi20093120909, Torrisi2010044705}, and brown coal~\cite{DANG201733} while being significantly lower than bituminous coal~\cite{Hu202012860}. This is likely due to the pore -H atoms being available to form interactions with CO$_2$. Similarly, the adsorption energies of CH$_4$ into functionalized carbophene systems are larger than those found for edge-functionalized graphene~\cite{Wood2012054702}, IRMOF-1~\cite{Arjmandi20181473}, brown coal~\cite{DANG201733}, and bituminous coal~\cite{Hu202012860}. Interestingly, the adsorption energies of H$_2$ into functionalized carbophenes were often less than the similarly studied edge-functionalized graphene~\cite{Ulman2014174708}, functionalized carbon nanotubes~\cite{SOLEYMANABADI2013115}. Compared to CH$_4$ and H$_2$, the preferential adsorption of CO$_2$ is likely due to the lone pairs on CO$_2$’s oxygen atoms. This suggests that N$_2$, the main component of air, will not foul carbophene-based air cleaners. Thus, these results demonstrate that functionalized carbophenes may be a suitable replacement for current greenhouse gas-capturing materials if their synthesis becomes reproducible and economical.

\begin{figure}
\centering
\includegraphics[clip,width=3 in, keepaspectratio]{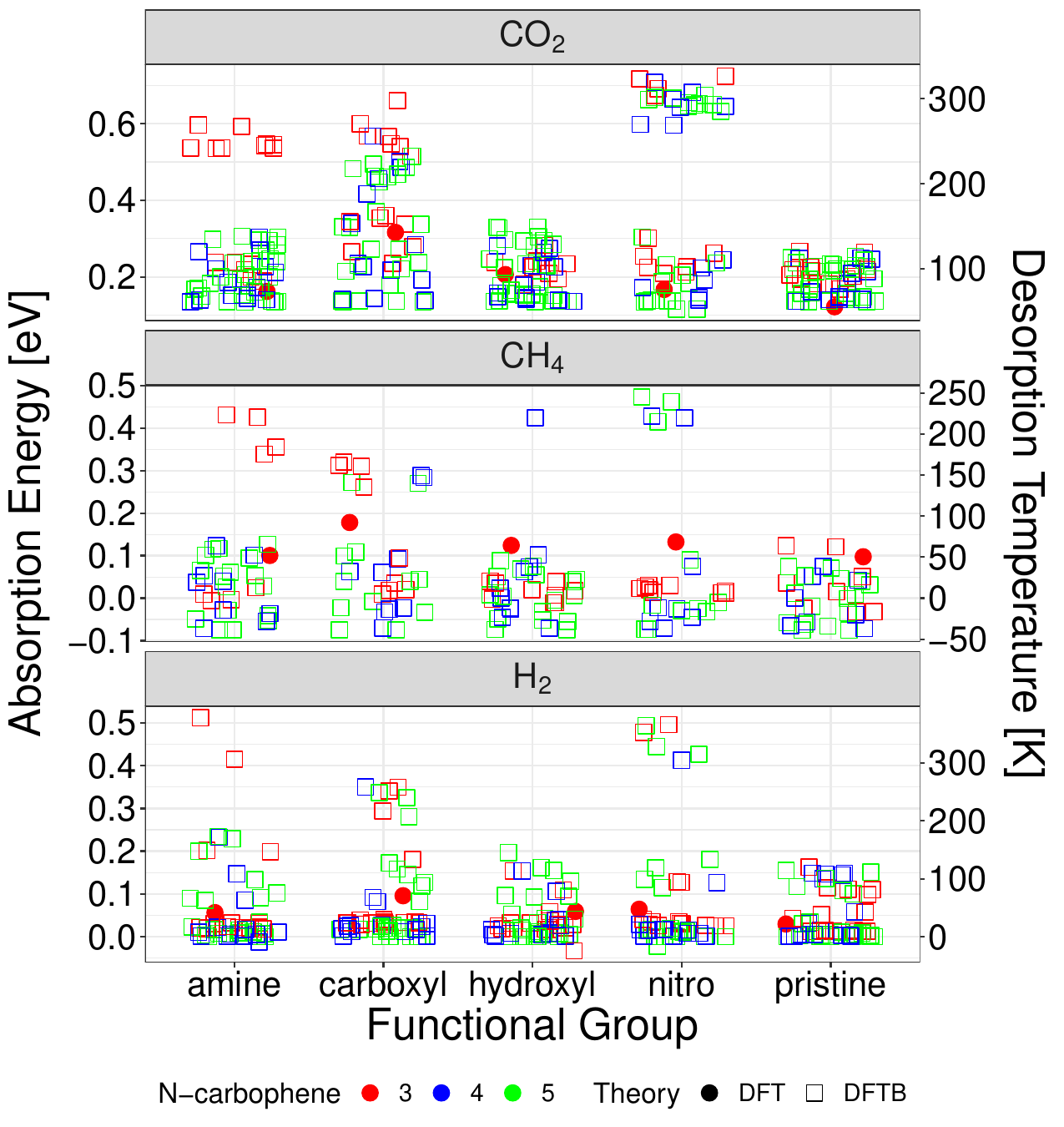}
\caption{Comparison of DFT and DFTB adsorption energies and desorption temperatures of CO$_2$, CH$_4$, or H$_2$ molecules into functionalized 3-, 4-, and 5-carbophenes.}
\label{fig:DFT_DFTB_comparison}
\end{figure}

The results and discussion above are based on a limited set of DFT calculations. A range of possible gas molecule adsorption configurations is needed to understand the possible gas molecule adsorption characteristics better. Computationally cost-effective DFTB calculations were performed for many more possible placements and orientations of adsorbate molecules (CO$_2$, CH$_4$, H$_2$) than were used in the DFT calculations. The DFTB simulations used many pseudo-random starting guess structures, the gas molecules placed in the center of the pore as used in the DFT-predicted structures, and expert-approved starting configurations for the gas molecules near adsorbing sites in the pores. Figure \ref{fig:DFT_DFTB_comparison} compares the DFT-based functionalized 3-carbophene adsorption results with DFTB-based adsorption results for functionalized 3-, 4-, and 5-carbophenes. These results demonstrate that the DFT and DFTB results give similar adsorption energies. Second, the adsorption energies are near zero even when the adsorbate is near a functional group but not close to an ideal position. For CH$_4$, adsorption energies for non-ideal positions can also be negative, demonstrating an endothermic reaction. Ideally placed and oriented adsorbates often have much higher adsorption energies than the same molecule that is shifted slightly. While we expect Coulombic interactions to play a role in the adsorptions, the configurations with the highest adsorption energies appear to have an apparent propensity for hydrogen bonding. These results demonstrate that organically functionalized carbophenes have adsorption energies usually seen in noble metal catalysts on reducible metal oxides. Thus, carbophenes may be an environmentally friendly solid-state method to capture and store gasses.

We can estimate the desorption temperature using the van't Hoff equation:
\begin{equation}
T_d = \frac{E_{ads}}{k_B} \left( \frac{ \Delta S}{R}  -  \ln{P}  \right)^{-1},
\end{equation}
where $E_{ads}$ is the adsorption energy, $k_B$ is the Boltzmann constant, $\Delta S$ is the change in entropy from a gas to a liquid, and R is the gas constant~\cite{ANIKINA2020100421}.  The $\Delta S$ for CO$_2$, CH$_4$, and H$_2$ are respectively 213.8 J/(mol$\cdot$K), 186.25 J/(mol$\cdot$K), and 130.680 J/(mol$\cdot$K) at 1 bar~\cite{Linstrom2017NISTwebbook}. Figure~\ref{fig:DFT_DFTB_comparison} presents the desorption temperature estimates for the gases we study.  The reported H$_2$ desorption temperatures fall within the US DOE's range (233 to 373 K) for hydrogen storage technologies~\cite{us2017doehydrogen}, making the functionalized carbophenes a good choice.  Whereas the desorption temperatures of CO$_2$ fall below the average temperature of flue gases, 350 K. Thus, carbophenes would have to be used in settings where the temperature of the gases is closer to ambient.  Thus, using functionalized carbophenes for CO$_2$ capture may need to be limited to air scrubbers in tight spaces~\cite{winton2016carbon}.

The gravimetric densities $\rho_G$ of the gasses in the carbophene-gas systems can be estimated by
\begin{equation}
\rho_G = \frac{\sum_{X_{\text{gas}}} N_X W_X} {\sum_{X_{\text{system}}} N_X W_X}
\end{equation}
where $W_X$ is the atomic weight of atom species $X$, $N_X$ is the number of $X$ atoms.  Using the 2-by-2 supercell models based on Figure~\ref{fig:LatVecSupercell}(e), the gravimetric densities of one adsorbed gas molecule are all $\rho_G \sim 0.0013$. The $\rho_G$ values found here are an order of magnitude smaller than the DOEHST targets for H$_2$ adsorption, but this study hasn't optimized the number of adsorption sites or the amount of adsorbed H$_2$ per pore. Further research into the maximum achievable gravimetric densities is warranted. Experimental research in porous carbons has demonstrated that pore sizes of $\sim 7$ Å provide the best CO$_2$ adsorptions near 3-carbophene pore sizes \cite{Adeniran2016994}.

\begin{figure}
\centering
\includegraphics[clip,width=3 in, keepaspectratio]{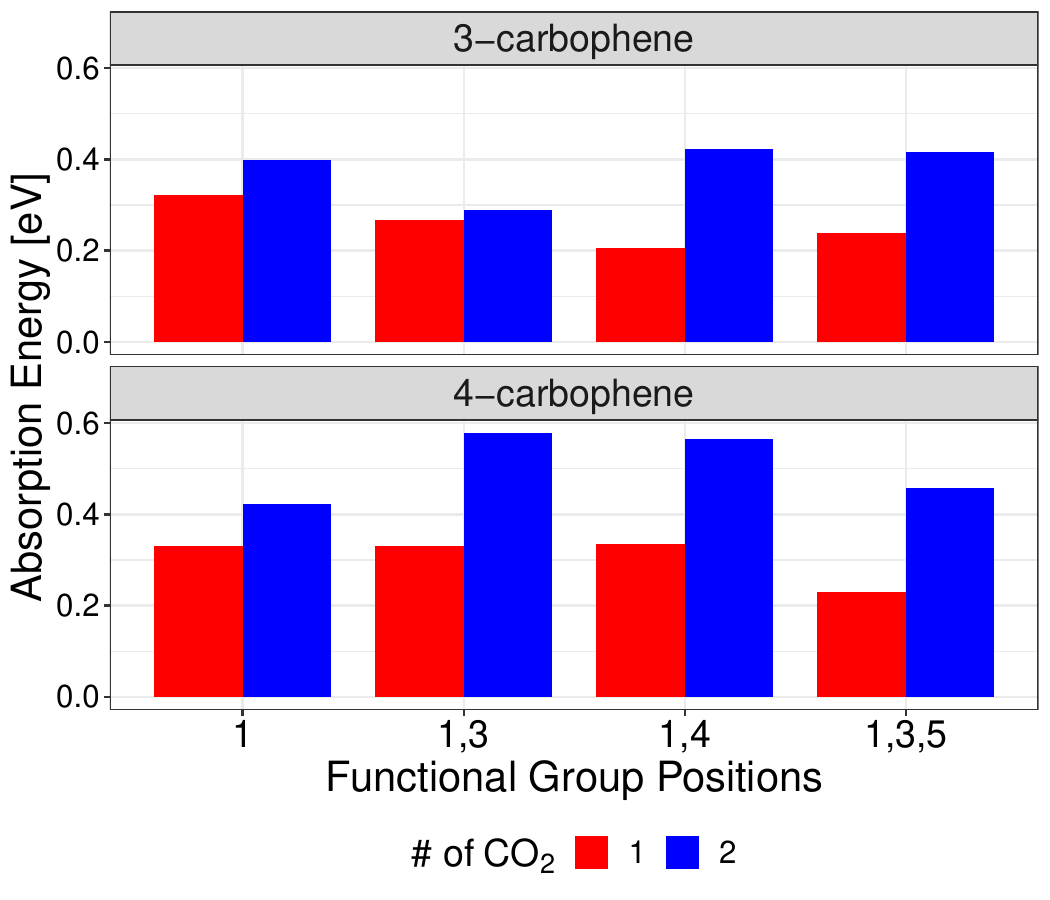}
\caption{adsorption energies of 1-carboxyl and 3-carboxyl structures with one or two adsorbed CO$_2$ molecules.}
\label{fig:JeanCharlesFigure1}
\end{figure}

Subsequently, we examined the adsorption energetics for multiple functionalizations of the carbophene pores to test whether increasing the density of functional groups can have a beneficial effect. Figure \ref{fig:JeanCharlesFigure1} presents the adsorption energies of one or two CO$_2$ molecules when one, two (in cis or trans configurations), or three carboxyl groups (in a 1,3,5 configuration) functionalize a carbophene pore. The difference in the 3- and 4-carbophene adsorption energies are negligible for the case where each has one carboxyl functional group. 4-carbophene has an adjacent -H atom further away than in 3-carbophene. Thus, the insignificant difference in their adsorption energies suggests that the -H atoms play a minor role in adsorption. Therefore, the critical interactions for the system's energy are all due to the carboxyl functional group.


\begin{figure}
\centering
\includegraphics[clip,width=3 in, keepaspectratio]{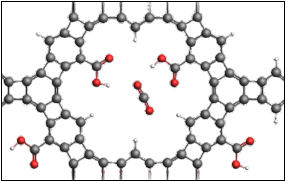}
\caption{Cis-2-carboxyl-3-carbophene unit cell with a CO$_2$ molecule adsorbate.}
\label{fig:JeanCharlesFigureA9}
\end{figure}

However, a change in adsorption energy is observed for cis-2-carboxyl functional group configurations. There is an increase in adsorption energy with 4-carbophene, which is not present for the trans functional group configuration. Figure~\ref{fig:JeanCharlesFigureA9} shows these functional groups can be close, leading to interactions that could diminish the adsorption energy. These interactions are not present, or are much weaker, in a trans configuration since functional groups find themselves on opposite sides of the carbophene pore.

With the addition of a second CO$_2$ molecule for the adsorbate, the increase in adsorption energy from 3-carbophene to 4-carbophene now becomes present for both cis and trans configurations. However, adding a second adsorbate leads to even more disruption of the adsorption, as the smaller pore for 3-carbophene leads to unwanted interactions. This behavior differs from that observed for 4-carbophene, which has much larger pores and where the adsorbate can interact almost exclusively with the functional groups. In addition, while for trans configuration, the interaction between adsorbed molecules was relatively weak, this is no longer the case with additional adsorbates, as two adsorbed CO$_2$ molecules in the same pore can interact with one another due to their proximity, altering the energetics of the system.

Considering two carboxyl groups in a trans configuration is energetically more favorable to adsorption in both cases, 1 or 2 CO$_2$ molecules. On the other hand, our results indicate that 4-carbophene could also favor CO$_2$ adsorption due to its large pores. A rough estimate of pore size differences indicates that 4-carbophene has approximately twice the pore size of 3-carbophene.

The addition of adsorbate molecules highlights the importance of the larger pore size of 4-carbophene. The adsorption energies are similar between the different structures with a single CO$_2$ molecule as an adsorbate. However, the incremental change from adding a second CO$_2$ was much more significant in the molecules with two functional groups. Molecules with more functional groups can more effectively handle the adsorbate. Thus, these structures are promising candidates for efficient adsorption of CO$_2$ molecules.

\begin{figure}
\centering
\includegraphics[clip,width=3 in, keepaspectratio]{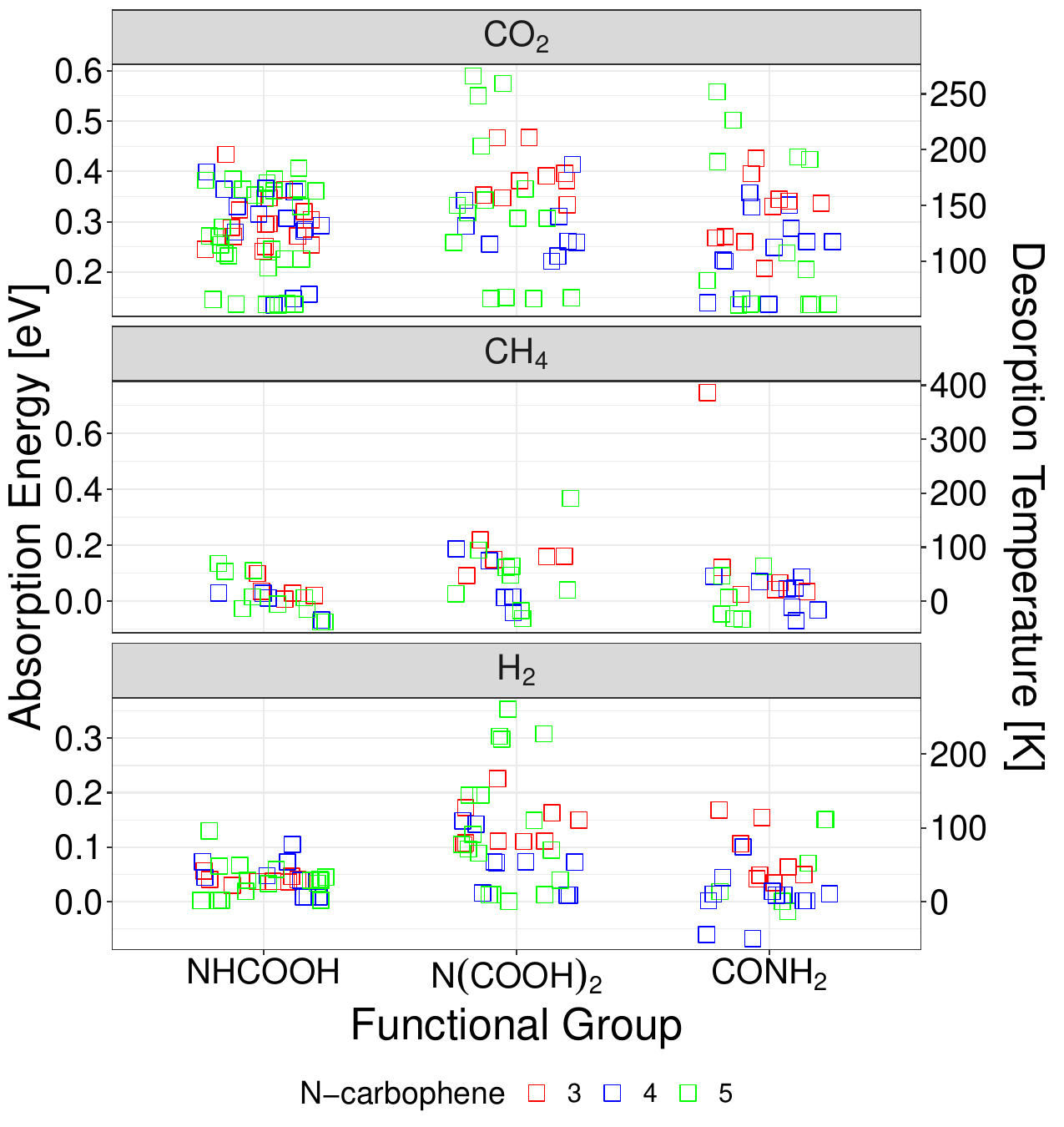}
\caption{Comparison of DFT and DFTB adsorption energies of CO$_2$, CH$_4$, or H$_2$ into 3-carbophene functionalized by CONH$_2$, N(COOH)$_2$, and NHCOOH.}
\label{fig:NHCOOH_N2COOH_comparison}
\end{figure}

The preceding analysis has focused on the adsorption effects of functional groups, resulting in adsorption energy of up to about 0.6 eV. To determine if we could push the adsorption energies to higher values using organic functional groups, we developed a few simple extended functionalizing molecules CONH$_2$, NHCOOH, and N(COOH)$_2$. While the adsorbate molecules (CO$_2$, CH$_4$, and H$_2$) were placed in positions and orientations that might be favorable to increased interaction with the functional groups, results do not demonstrate an increase in the adsorption energies; see Figure \ref{fig:NHCOOH_N2COOH_comparison}.

\section*{Conclusion}

In conclusion, functionalized carbophenes may be used in gas capture and separation due to their favorable adsorption energies, large surface area, and high pore density. This work demonstrates that organo-functionalized carbophenes have the propensity to adsorb molecular hydrogen at energies usually associated with metal-containing materials. Further, the estimated H$_2$ desorption temperatures align with US DOE targets for hydrogen storage technologies. Functionalized carbophenes have CO$_2$ adsorption energies higher than many of the leading candidates for next-generation CO$_2$ reduction materials but still have desorption temperature estimates that are well below flue exhaust temperatures. Adding more than one functional group to 4-carbophene significantly increases adsorption potential, as 4-carbophene's significantly larger pore size allows for more gas molecules. Future avenues of this study would include calculating these systems using thermodynamics to determine this material's diffusivity and membrane properties. Further investigation should include multi-layered systems to better understand this material's adsorption properties.

\section*{Acknowledgement}

C.E.J., K.L., R.M.A., and V.A.D. were supported by the United States funding agency NSF under award \#2113011, and advanced computing resources from University of Hawaii Information Technology Services – Cyberinfrastructure, funded in part by the National Science Foundation awards \#2201428 and \#2232862. J.Y. was supported by the United States funding agency NSF (Grant Number DMR-2121848) University of Hawaii–University of Washington PREM.  E.L., J.-C.M., and G.P. acknowledge Compute Canada’s high performance computing resources (Graham cluster located at the University of Waterloo) via the Shared Hierarchical Academic Research Computing Network (SHARCNET) consortium. J.K. was supported by Hawai‘i Space Grant Consortium NASA Space Grant Undergraduate Traineeship.  R. Paupitz acknowledges Brazilian funding agencies: FAPESP (grant \#2018/03961-5) and CNPq (grants \#437034/2018-6, \#310369/2017-7) and computing resources supplied by the "Centro Nacional de Processamento de Alto Desempenho em São Paulo
(CENAPAD-SP)" and the National Laboratory for Scientific Computing (LNCC).


\section*{{CRediT} authorship contribution statement}

Chad E. Junkermeier: Conceptualization, Data curation, Investigation, Methodology, Project administration, Resources, Software, Validation, Visualization, Writing – original draft, Writing – review \& editing.
Evan Larmand: Investigation, Writing – original draft.
Jean-Charles Morais: Investigation, Writing – original draft.
Jedediah Kobebel: Investigation, Writing – review \& editing.
Kat Lavarez: Investigation, Writing – review \& editing.
Ralph Martin Adra: Investigation, Writing – review \& editing.
Jirui Yang: Investigation, Writing – review \& editing.
Valeria Aparicio Diaz: Investigation, Writing – review \& editing.
George Psofogiannakis: Conceptualization, Resources, Writing – review \& editing.
Ricardo Paupitz: Resources, Writing – review \& editing.

\section{Data Availability}

The raw data required to reproduce these findings are available to download from Mendeley Data \cite{gasstoragecarbophenes}.




\bibliographystyle{elsarticle-num}
\bibliography{carbophene}

\begin{thebibliography}{10}
\expandafter\ifx\csname url\endcsname\relax
  \def\url#1{\texttt{#1}}\fi
\expandafter\ifx\csname urlprefix\endcsname\relax\def\urlprefix{URL }\fi
\expandafter\ifx\csname href\endcsname\relax
  \def\href#1#2{#2} \def\path#1{#1}\fi

\bibitem{stocker2014climate}
T.~Stocker, Climate change 2013: the physical science basis: Working Group I
  contribution to the Fifth assessment report of the Intergovernmental Panel on
  Climate Change, Cambridge university press, 2014.

\bibitem{Zhao17}
G.~Ji, M.~Zhao, \href{https://doi.org/10.5772/65723}{Membrane separation
  technology in carbon capture}, in: Y.~Yun (Ed.), Recent Advances in Carbon
  Capture and Storage, IntechOpen, Rijeka, 2017, Ch.~3, pp. 59--90.
\newblock \href {http://dx.doi.org/10.5772/65723} {\path{doi:10.5772/65723}}.
\newline\urlprefix\url{https://doi.org/10.5772/65723}

\bibitem{winton2016carbon}
D.~Winton, J.~Isobe, P.~Henson, A.~MacKnight, S.~Yates, D.~Schuck,
  \href{https://ttu-ir.tdl.org/handle/2346/67727}{Carbon dioxide removal
  technologies for space vehicles-past, present, and future}, in: 46th
  International Conference on Environmental Systems, 2016, pp. 1--10.
\newblock \href
  {http://arxiv.org/abs/https://ttu-ir.tdl.org/bitstream/handle/2346/67727/ICES_2016_425.pdf}
  {\path{arXiv:https://ttu-ir.tdl.org/bitstream/handle/2346/67727/ICES_2016_425.pdf}}.
\newline\urlprefix\url{https://ttu-ir.tdl.org/handle/2346/67727}

\bibitem{Zilberman20154}
P.~Zilberman, \href{https://doi.org/10.1515/amma-2015-0023}{The co absorber
  based on lioh}, Acta Marisiensis - Seria Medica 61~(1) (2015) 4--6.
\newblock \href {http://dx.doi.org/doi:10.1515/amma-2015-0023}
  {\path{doi:doi:10.1515/amma-2015-0023}}.
\newline\urlprefix\url{https://doi.org/10.1515/amma-2015-0023}

\bibitem{cmarik2018co}
G.~E. Cmarik, J.~C. Knox,
  \href{https://ntrs.nasa.gov/citations/20180006333}{Co-adsorption of carbon
  dioxide on zeolite 13x in the presence of preloaded water}, in: International
  Conference on Environmental Systems, no. ICES-2018-3 in 1, 2018, pp. 1--10.
\newblock \href
  {http://arxiv.org/abs/https://ntrs.nasa.gov/api/citations/20180006333/downloads/20180006333.pdf}
  {\path{arXiv:https://ntrs.nasa.gov/api/citations/20180006333/downloads/20180006333.pdf}}.
\newline\urlprefix\url{https://ntrs.nasa.gov/citations/20180006333}

\bibitem{D0CS00049C}
R.-R. Liang, S.-Y. Jiang, R.-H. A, X.~Zhao,
  \href{http://dx.doi.org/10.1039/D0CS00049C}{Two-dimensional covalent organic
  frameworks with hierarchical porosity}, Chem. Soc. Rev. 49 (2020) 3920--3951.
\newblock \href {http://dx.doi.org/10.1039/D0CS00049C}
  {\path{doi:10.1039/D0CS00049C}}.
\newline\urlprefix\url{http://dx.doi.org/10.1039/D0CS00049C}

\bibitem{Lim5b12090}
G.~Lim, K.~B. Lee, H.~C. Ham,
  \href{https://doi.org/10.1021/acs.jpcc.5b12090}{Effect of n-containing
  functional groups on co2 adsorption of carbonaceous materials: A density
  functional theory approach}, The Journal of Physical Chemistry C 120~(15)
  (2016) 8087--8095.
\newblock \href {http://arxiv.org/abs/https://doi.org/10.1021/acs.jpcc.5b12090}
  {\path{arXiv:https://doi.org/10.1021/acs.jpcc.5b12090}}, \href
  {http://dx.doi.org/10.1021/acs.jpcc.5b12090}
  {\path{doi:10.1021/acs.jpcc.5b12090}}.
\newline\urlprefix\url{https://doi.org/10.1021/acs.jpcc.5b12090}

\bibitem{Arjmandi20181473}
M.~Arjmandi, M.~Pourafshari~Chenar, M.~Peyravi, M.~Jahanshahi, A.~Arjmandi,
  A.~Shokuhi~Rad, \href{https://www.ije.ir/article_73300.html}{Interpreting the
  {CO$_2$} adsorption on functionalized organic group of {IRMOF-1}: A {B3LYP
  DFT} based study}, International Journal of Engineering 31~(9) (2018)
  1473--1479.
\newblock \href
  {http://arxiv.org/abs/https://www.ije.ir/article_73300_68af936ad128b995e9f7e66ecc06b70c.pdf}
  {\path{arXiv:https://www.ije.ir/article_73300_68af936ad128b995e9f7e66ecc06b70c.pdf}}.
\newline\urlprefix\url{https://www.ije.ir/article_73300.html}

\bibitem{Torrisi20093120909}
A.~Torrisi, C.~Mellot-Draznieks, R.~G. Bell,
  \href{https://doi.org/10.1063/1.3120909}{Impact of ligands on co2 adsorption
  in metal-organic frameworks: First principles study of the interaction of co2
  with functionalized benzenes. i. inductive effects on the aromatic ring}, The
  Journal of Chemical Physics 130~(19) (2009) 194703.
\newblock \href {http://arxiv.org/abs/https://doi.org/10.1063/1.3120909}
  {\path{arXiv:https://doi.org/10.1063/1.3120909}}, \href
  {http://dx.doi.org/10.1063/1.3120909} {\path{doi:10.1063/1.3120909}}.
\newline\urlprefix\url{https://doi.org/10.1063/1.3120909}

\bibitem{Torrisi2010044705}
A.~Torrisi, C.~Mellot-Draznieks, R.~G. Bell,
  \href{https://doi.org/10.1063/1.3276105}{Impact of ligands on co2 adsorption
  in metal-organic frameworks: First principles study of the interaction of co2
  with functionalized benzenes. ii. effect of polar and acidic substituents},
  The Journal of Chemical Physics 132~(4) (2010) 044705.
\newblock \href {http://arxiv.org/abs/https://doi.org/10.1063/1.3276105}
  {\path{arXiv:https://doi.org/10.1063/1.3276105}}, \href
  {http://dx.doi.org/10.1063/1.3276105} {\path{doi:10.1063/1.3276105}}.
\newline\urlprefix\url{https://doi.org/10.1063/1.3276105}

\bibitem{SOLEYMANABADI2013115}
H.~Soleymanabadi, J.~Kakemam,
  \href{https://www.sciencedirect.com/science/article/pii/S1386947713002245}{A
  dft study of h2 adsorption on functionalized carbon nanotubes}, Physica E:
  Low-dimensional Systems and Nanostructures 54 (2013) 115--117.
\newblock \href {http://dx.doi.org/https://doi.org/10.1016/j.physe.2013.06.015}
  {\path{doi:https://doi.org/10.1016/j.physe.2013.06.015}}.
\newline\urlprefix\url{https://www.sciencedirect.com/science/article/pii/S1386947713002245}

\bibitem{HIRSCHER2020153548}
M.~Hirscher, V.~A. Yartys, M.~Baricco, J.~{Bellosta von Colbe}, D.~Blanchard,
  R.~C. Bowman, D.~P. Broom, C.~E. Buckley, F.~Chang, P.~Chen, Y.~W. Cho, J.-C.
  Crivello, F.~Cuevas, W.~I. David, P.~E. {de Jongh}, R.~V. Denys, M.~Dornheim,
  M.~Felderhoff, Y.~Filinchuk, G.~E. Froudakis, D.~M. Grant, E.~M. Gray, B.~C.
  Hauback, T.~He, T.~D. Humphries, T.~R. Jensen, S.~Kim, Y.~Kojima,
  M.~Latroche, H.-W. Li, M.~V. Lototskyy, J.~W. Makepeace, K.~T. Møller,
  L.~Naheed, P.~Ngene, D.~Noréus, M.~M. Nygård, S.~ichi Orimo,
  M.~Paskevicius, L.~Pasquini, D.~B. Ravnsbæk, M.~{Veronica Sofianos}, T.~J.
  Udovic, T.~Vegge, G.~S. Walker, C.~J. Webb, C.~Weidenthaler, C.~Zlotea,
  \href{https://www.sciencedirect.com/science/article/pii/S0925838819347942}{Materials
  for hydrogen-based energy storage – past, recent progress and future
  outlook}, Journal of Alloys and Compounds 827 (2020) 153548.
\newblock \href
  {http://dx.doi.org/https://doi.org/10.1016/j.jallcom.2019.153548}
  {\path{doi:https://doi.org/10.1016/j.jallcom.2019.153548}}.
\newline\urlprefix\url{https://www.sciencedirect.com/science/article/pii/S0925838819347942}

\bibitem{SIMANULLANG202229808}
M.~Simanullang, L.~Prost,
  \href{https://www.sciencedirect.com/science/article/pii/S0360319922029597}{Nanomaterials
  for on-board solid-state hydrogen storage applications}, International
  Journal of Hydrogen Energy 47~(69) (2022) 29808--29846.
\newblock \href
  {http://dx.doi.org/https://doi.org/10.1016/j.ijhydene.2022.06.301}
  {\path{doi:https://doi.org/10.1016/j.ijhydene.2022.06.301}}.
\newline\urlprefix\url{https://www.sciencedirect.com/science/article/pii/S0360319922029597}

\bibitem{dillon2005nrel}
\href{https://www.hydrogen.energy.gov/pdfs/review05/stp_63_heben.pdf}{NREL
  activities in DOE Carbon-based Materials Center of Excellence}.
\newline\urlprefix\url{https://www.hydrogen.energy.gov/pdfs/review05/stp_63_heben.pdf}

\bibitem{D2SC00871H}
K.~Shun, K.~Mori, S.~Masuda, N.~Hashimoto, Y.~Hinuma, H.~Kobayashi,
  H.~Yamashita, \href{http://dx.doi.org/10.1039/D2SC00871H}{Revealing hydrogen
  spillover pathways in reducible metal oxides}, Chem. Sci. 13 (2022)
  8137--8147.
\newblock \href {http://dx.doi.org/10.1039/D2SC00871H}
  {\path{doi:10.1039/D2SC00871H}}.
\newline\urlprefix\url{http://dx.doi.org/10.1039/D2SC00871H}

\bibitem{us2017doehydrogen}
{US Department of Energy},
  \href{https://www.energy.gov/eere/fuelcells/doe-technical-targets-onboard-hydrogen-storage-light-duty-vehicles
  (Accessed 4 July 2023)}{{DOE} technical targets for onboard hydrogen storage
  for light-duty vehicles}, Tech. rep., Department of Energy, accessed 4 July
  2023 (2017).
\newline\urlprefix\url{https://www.energy.gov/eere/fuelcells/doe-technical-targets-onboard-hydrogen-storage-light-duty-vehicles
  (Accessed 4 July 2023)}

\bibitem{du1740796}
Q.-S. Du, P.-D. Tang, H.-L. Huang, F.-L. Du, K.~Huang, N.-Z. Xie, S.-Y. Long,
  Y.-M. Li, J.-S. Qiu, R.-B. Huang, \href{https://doi.org/10.1038/srep40796}{A
  new type of two-dimensional carbon crystal prepared from
  1,3,5-trihydroxybenzene}, Scientific Reports 7 (2017) 40796.
\newblock \href {http://dx.doi.org/10.1038/srep40796}
  {\path{doi:10.1038/srep40796}}.
\newline\urlprefix\url{https://doi.org/10.1038/srep40796}

\bibitem{junkermeier2019simplecarbophene}
C.~E. Junkermeier, J.~P. Luben, R.~Paupitz,
  \href{https://doi.org/10.1088/2F2053-1591/2Fab4513}{N-carbophenes:
  two-dimensional covalent organic frameworks derived from linear
  n-phenylenes}, Materials Research Express 6~(11) (2019) 115103.
\newblock \href {http://dx.doi.org/10.1088/2053-1591/ab4513}
  {\path{doi:10.1088/2053-1591/ab4513}}.
\newline\urlprefix\url{https://doi.org/10.1088/2F2053-1591/2Fab4513}

\bibitem{Junkermeier2022covalent}
C.~E. Junkermeier, G.~Psofogiannakis, R.~Paupitz,
  \href{https://doi.org/10.1088/2053-1591/ac4c19}{Covalent adsorption of
  functional groups on n-carbophenes}, Materials Research Express\href
  {http://dx.doi.org/10.1088/2053-1591/ac4c19}
  {\path{doi:10.1088/2053-1591/ac4c19}}.
\newline\urlprefix\url{https://doi.org/10.1088/2053-1591/ac4c19}

\bibitem{Perdew2008136406}
J.~P. Perdew, A.~Ruzsinszky, G.~I. Csonka, O.~A. Vydrov, G.~E. Scuseria, L.~A.
  Constantin, X.~Zhou, K.~Burke,
  \href{https://link.aps.org/doi/10.1103/PhysRevLett.100.136406}{Restoring the
  density-gradient expansion for exchange in solids and surfaces}, Phys. Rev.
  Lett. 100 (2008) 136406.
\newblock \href {http://dx.doi.org/10.1103/PhysRevLett.100.136406}
  {\path{doi:10.1103/PhysRevLett.100.136406}}.
\newline\urlprefix\url{https://link.aps.org/doi/10.1103/PhysRevLett.100.136406}

\bibitem{PhysRevB.44.7888}
G.~te~Velde, E.~J. Baerends,
  \href{https://link.aps.org/doi/10.1103/PhysRevB.44.7888}{Precise
  density-functional method for periodic structures}, Phys. Rev. B 44 (1991)
  7888--7903.
\newblock \href {http://dx.doi.org/10.1103/PhysRevB.44.7888}
  {\path{doi:10.1103/PhysRevB.44.7888}}.
\newline\urlprefix\url{https://link.aps.org/doi/10.1103/PhysRevB.44.7888}

\bibitem{GrimmeD3}
S.~Grimme, J.~Antony, S.~Ehrlich, H.~Krieg,
  \href{https://doi.org/10.1063/1.3382344}{A consistent and accurate ab initio
  parametrization of density functional dispersion correction (dft-d) for the
  94 elements h-pu}, The Journal of Chemical Physics 132~(15) (2010) 154104.
\newblock \href {http://arxiv.org/abs/https://doi.org/10.1063/1.3382344}
  {\path{arXiv:https://doi.org/10.1063/1.3382344}}, \href
  {http://dx.doi.org/10.1063/1.3382344} {\path{doi:10.1063/1.3382344}}.
\newline\urlprefix\url{https://doi.org/10.1063/1.3382344}

\bibitem{Elstner1998}
M.~Elstner, D.~Porezag, G.~Jungnickel, J.~Elsner, M.~Haugk, T.~Frauenheim,
  S.~Suhai, G.~Seifert,
  \href{http://link.aps.org/doi/10.1103/PhysRevB.58.7260}{{Self-consistent-charge
  density-functional tight-binding method for simulations of complex materials
  properties}}, Physical Review B 58~(11) (1998) 7260--7268.
\newblock \href {http://dx.doi.org/10.1103/PhysRevB.58.7260}
  {\path{doi:10.1103/PhysRevB.58.7260}}.
\newline\urlprefix\url{http://link.aps.org/doi/10.1103/PhysRevB.58.7260}

\bibitem{aradi2007dftb}
B.~Aradi, B.~Hourahine, T.~Frauenheim,
  \href{https://doi.org/10.1021/jp070186p}{{DFTB}+, a sparse matrix-based
  implementation of the {DFTB} method}, The Journal of Physical Chemistry A
  111~(26) (2007) 5678--5684, pMID: 17567110.
\newblock \href {http://arxiv.org/abs/https://doi.org/10.1021/jp070186p}
  {\path{arXiv:https://doi.org/10.1021/jp070186p}}, \href
  {http://dx.doi.org/10.1021/jp070186p} {\path{doi:10.1021/jp070186p}}.
\newline\urlprefix\url{https://doi.org/10.1021/jp070186p}

\bibitem{manzano2012}
H.~Manzano, A.~N. Enyashin, J.~S. Dolado, A.~Ayuela, J.~Frenzel, G.~Seifert,
  \href{http://dx.doi.org/10.1002/adma.201103704}{Do cement nanotubes exist?},
  Advanced Materials 24~(24) (2012) 3239--3245.
\newblock \href {http://dx.doi.org/10.1002/adma.201103704}
  {\path{doi:10.1002/adma.201103704}}.
\newline\urlprefix\url{http://dx.doi.org/10.1002/adma.201103704}

\bibitem{frenzel2004semi}
J.~Frenzel, A.~Oliveira, N.~Jardillier, T.~Heine, G.~Seifert,
  Semi-relativistic, self-consistent charge slater-koster tables for
  density-functional based tight-binding ({DFTB}) for materials science
  simulations, Zeolites 2~(3) (2004) 7.
\newblock \href {http://dx.doi.org/10.3762/bjnano.1.8}
  {\path{doi:10.3762/bjnano.1.8}}.

\bibitem{lukose2010reticular}
B.~Lukose, A.~Kuc, J.~Frenzel, T.~Heine, On the reticular construction concept
  of covalent organic frameworks, Beilstein journal of nanotechnology 1 (2010)
  60.
\newblock \href {http://dx.doi.org/10.3762/bjnano.1.8}
  {\path{doi:10.3762/bjnano.1.8}}.

\bibitem{Rezac20174804}
J.~Řezáč, \href{https://doi.org/10.1021/acs.jctc.7b00629}{Empirical
  self-consistent correction for the description of hydrogen bonds in dftb3},
  Journal of Chemical Theory and Computation 13~(10) (2017) 4804--4817, pMID:
  28949517.
\newblock \href {http://arxiv.org/abs/https://doi.org/10.1021/acs.jctc.7b00629}
  {\path{arXiv:https://doi.org/10.1021/acs.jctc.7b00629}}, \href
  {http://dx.doi.org/10.1021/acs.jctc.7b00629}
  {\path{doi:10.1021/acs.jctc.7b00629}}.
\newline\urlprefix\url{https://doi.org/10.1021/acs.jctc.7b00629}

\bibitem{gasstoragecarbophenes}
C.~Junkermeier, J.~Kobobel, K.~Lavarez, M.~Adra, V.~Diaz, J.~Yang,
  G.~Psofogiannakis, R.~Paupitz,
  \href{https://data.mendeley.com/datasets/bxkbbs2553/1}{Dftb based models of
  gas molecules adsorbed in functionalized carbophenes} (2023).
\newblock \href {http://dx.doi.org/10.17632/BXKBBS2553.1}
  {\path{doi:10.17632/BXKBBS2553.1}}.
\newline\urlprefix\url{https://data.mendeley.com/datasets/bxkbbs2553/1}

\bibitem{Solenov13115502}
D.~Solenov, C.~Junkermeier, T.~L. Reinecke, K.~A. Velizhanin,
  \href{http://link.aps.org/doi/10.1103/PhysRevLett.111.115502}{Tunable
  adsorbate-adsorbate interactions on graphene}, Phys. Rev. Lett. 111 (2013)
  115502.
\newblock \href {http://dx.doi.org/10.1103/PhysRevLett.111.115502}
  {\path{doi:10.1103/PhysRevLett.111.115502}}.
\newline\urlprefix\url{http://link.aps.org/doi/10.1103/PhysRevLett.111.115502}

\bibitem{Li2017624}
P.~Li, N.~A. Vermeulen, C.~D. Malliakas, D.~A. Gómez-Gualdrón, A.~J. Howarth,
  B.~L. Mehdi, A.~Dohnalkova, N.~D. Browning, M.~O’Keeffe, O.~K. Farha,
  \href{https://www.science.org/doi/abs/10.1126/science.aam7851}{Bottom-up
  construction of a superstructure in a porous uranium-organic crystal},
  Science 356~(6338) (2017) 624--627.
\newblock \href
  {http://arxiv.org/abs/https://www.science.org/doi/pdf/10.1126/science.aam7851}
  {\path{arXiv:https://www.science.org/doi/pdf/10.1126/science.aam7851}}, \href
  {http://dx.doi.org/10.1126/science.aam7851}
  {\path{doi:10.1126/science.aam7851}}.
\newline\urlprefix\url{https://www.science.org/doi/abs/10.1126/science.aam7851}

\bibitem{DANG201733}
Y.~Dang, L.~Zhao, X.~Lu, J.~Xu, P.~Sang, S.~Guo, H.~Zhu, W.~Guo,
  \href{https://www.sciencedirect.com/science/article/pii/S0169433217317981}{Molecular
  simulation of co2/ch4 adsorption in brown coal: Effect of oxygen-, nitrogen-,
  and sulfur-containing functional groups}, Applied Surface Science 423 (2017)
  33--42.
\newblock \href
  {http://dx.doi.org/https://doi.org/10.1016/j.apsusc.2017.06.143}
  {\path{doi:https://doi.org/10.1016/j.apsusc.2017.06.143}}.
\newline\urlprefix\url{https://www.sciencedirect.com/science/article/pii/S0169433217317981}

\bibitem{Hu202012860}
Y.~Hu, S.~Wang, Y.~He,
  \href{https://doi.org/10.1021/acs.energyfuels.0c02497}{Investigation of the
  coal oxidation effect on competitive adsorption characteristics of co2/ch4},
  Energy \& Fuels 34~(10) (2020) 12860--12869.
\newblock \href
  {http://arxiv.org/abs/https://doi.org/10.1021/acs.energyfuels.0c02497}
  {\path{arXiv:https://doi.org/10.1021/acs.energyfuels.0c02497}}, \href
  {http://dx.doi.org/10.1021/acs.energyfuels.0c02497}
  {\path{doi:10.1021/acs.energyfuels.0c02497}}.
\newline\urlprefix\url{https://doi.org/10.1021/acs.energyfuels.0c02497}

\bibitem{Wood2012054702}
B.~C. Wood, S.~Y. Bhide, D.~Dutta, V.~S. Kandagal, A.~D. Pathak, S.~N.
  Punnathanam, K.~G. Ayappa, S.~Narasimhan,
  \href{https://doi.org/10.1063/1.4736568}{Methane and carbon dioxide
  adsorption on edge-functionalized graphene: A comparative dft study}, The
  Journal of Chemical Physics 137~(5) (2012) 054702.
\newblock \href {http://arxiv.org/abs/https://doi.org/10.1063/1.4736568}
  {\path{arXiv:https://doi.org/10.1063/1.4736568}}, \href
  {http://dx.doi.org/10.1063/1.4736568} {\path{doi:10.1063/1.4736568}}.
\newline\urlprefix\url{https://doi.org/10.1063/1.4736568}

\bibitem{Ulman2014174708}
K.~Ulman, D.~Bhaumik, B.~C. Wood, S.~Narasimhan,
  \href{https://doi.org/10.1063/1.4873435}{Physical origins of weak h2 binding
  on carbon nanostructures: Insight from ab initio studies of chemically
  functionalized graphene nanoribbons}, The Journal of Chemical Physics
  140~(17) (2014) 174708.
\newblock \href {http://arxiv.org/abs/https://doi.org/10.1063/1.4873435}
  {\path{arXiv:https://doi.org/10.1063/1.4873435}}, \href
  {http://dx.doi.org/10.1063/1.4873435} {\path{doi:10.1063/1.4873435}}.
\newline\urlprefix\url{https://doi.org/10.1063/1.4873435}

\bibitem{ANIKINA2020100421}
E.~Anikina, A.~Banerjee, V.~Beskachko, R.~Ahuja,
  \href{https://www.sciencedirect.com/science/article/pii/S246860692030040X}{Influence
  of kubas-type interaction of b–ni codoped graphdiyne with hydrogen
  molecules on desorption temperature and storage efficiency}, Materials Today
  Energy 16 (2020) 100421.
\newblock \href
  {http://dx.doi.org/https://doi.org/10.1016/j.mtener.2020.100421}
  {\path{doi:https://doi.org/10.1016/j.mtener.2020.100421}}.
\newline\urlprefix\url{https://www.sciencedirect.com/science/article/pii/S246860692030040X}

\bibitem{Linstrom2017NISTwebbook}
P.~Linstrom, Nist chemistry webbook - srd 69, Tech. rep., National Institute of
  Standards and Technology, Gaithersburg, MD, dOI: 10.1021/je000236i (accessed
  July 8, 2023). (2017).

\bibitem{Adeniran2016994}
B.~Adeniran, R.~Mokaya, \href{https://doi.org/10.1021/acs.chemmater.5b05020}{Is
  n-doping in porous carbons beneficial for co2 storage? experimental
  demonstration of the relative effects of pore size and n-doping}, Chemistry
  of Materials 28~(3) (2016) 994--1001.
\newblock \href
  {http://arxiv.org/abs/https://doi.org/10.1021/acs.chemmater.5b05020}
  {\path{arXiv:https://doi.org/10.1021/acs.chemmater.5b05020}}, \href
  {http://dx.doi.org/10.1021/acs.chemmater.5b05020}
  {\path{doi:10.1021/acs.chemmater.5b05020}}.
\newline\urlprefix\url{https://doi.org/10.1021/acs.chemmater.5b05020}

\end{thebibliography}







\end{document}


\begin{frontmatter}



\title{Ab initio investigation of carbon capture in carbophenes}

\author[label1,label2]{Chad E. Junkermeier}
\author[label3]{Evan Larmand}
\author[label3]{Jean-Charles Morais}
\author[label4]{Jedediah Kobebel}
\author[label5]{Ricardo Paupitz}
\author[label3]{George Psofogiannakis}

\address[label1]{Department of Physics and Astronomy, University of Hawai`i at Mānoa, Honolulu HI 96822, USA}
\address[label2]{Materials Computation Laboratory, University of Hawai`i Maui College, Kahului HI 96732, USA}
\address[label3]{Department of Chemical and Biological Engineering, University of Ottawa, Ottawa, Ontario, Canada}
\address[label4]{Science, Technology, Engineering, and Mathematics Department, University of Hawai`i Maui College, Kahului HI 96732, USA}

\address[label5]{Departamento de F\'{\i}sica, IGCE, Universidade Estadual Paulista, UNESP, 13506-900, Rio Claro, SP, Brazil}

\end{frontmatter}


\beginsupplement
\section*{Gravimetric Densities}

\begin{table*}[!htb]
    \centering
\caption{Gravimetic densities of 2-by-2 supercells as computed from Eqn. 4 of the text.}
\begin{tabular}{lccc}
\toprule
\multirow{2}{1in}{Carbophene functional group} & \multicolumn{3}{c}{Adsorbed gas molecule} \\
\cmidrule{2-4} 
	&	CO$_2$	&	CH$_4$	&	H$_2$	\\
\midrule
COOH	&	0.001297643	&	0.001321432	&	0.001333695	\\
NH$_2$	&	0.001322323	&	0.001347035	&	0.001359779	\\
NO$_2$	&	0.001296818	&	0.001320577	&	0.001332824	\\
OH	&	0.001358877	&	0.001346149	&	0.001358877	\\
Pristine	&	0.001335476	&	0.001360686	&	0.001373692	\\
\bottomrule
    \end{tabular}
    \label{table:chadtable1}
\end{table*}






